# Comments on the claimed observation of the Wigner-Huntington Transition to Metallic Hydrogen

M.I. Eremets, A. P. Drozdov

Max-Planck-Institut fur Chemie, Hahn-Meitner Weg 1, 55128 Mainz, Germany

In their recent work Dias and Silvera (*Science* 2017) claim to have observed the Wigner-Huntington transition of hydrogen to a metallic state (MH) at a pressure of 495 GPa at low temperatures. The evidence for this transition is based on a high electron carrier density deduced from a Drude free electron model fitted to the reflectivity of the sample. Based on our analysis of the reflectivity data we find no convincing evidence for metallic hydrogen in their published data. The pressure determination is also ambiguous – it should be ~630 GPa according to the presented Raman spectrum.
For comparison, we present our own data on the observation of highly reflecting hydrogen at pressures of 350-400 GPa. The appearance of metallic reflectivity is accompanied with a finite electrical conductivity of the sample. We argue that the actual pressure in the experiment of Dias and Silvera is likely below 400 GPa. In this case the observed enhanced reflectivity would be related to the phase transformation to conductive state published in arXiv:1601.04479.

We start by analyzing Dias and Silvera (DS) pressure measurements. Pressures up to ~300 GPa were determined from the known shift of hydrogen vibron in IR spectra (Zha, Liu et al. 2012). Pressure in this range was likely overestimated as the raw, not fitted, IR peaks at pressure 338 GPa correspond to pressure of 305 GPa according to Ref (Zha, Liu et al. 2012). Higher pressures were determined from the load applied to the diamond anvil cell (DAC). This empirical method is unacceptable as pressure in the sample depends strongly on the particular geometry of the anvils, gasket etc. and varies from run to run (Fig. 1). Moreover, DS use a linear pressure/load dependence (Fig. 1a), which is not realistic – it should be sublinear as the average pressure increases with load (Fig.1).

The highest pressure, where the sample started to reflect, was determined by the seemly more reliable Raman spectrum of the stressed diamond.

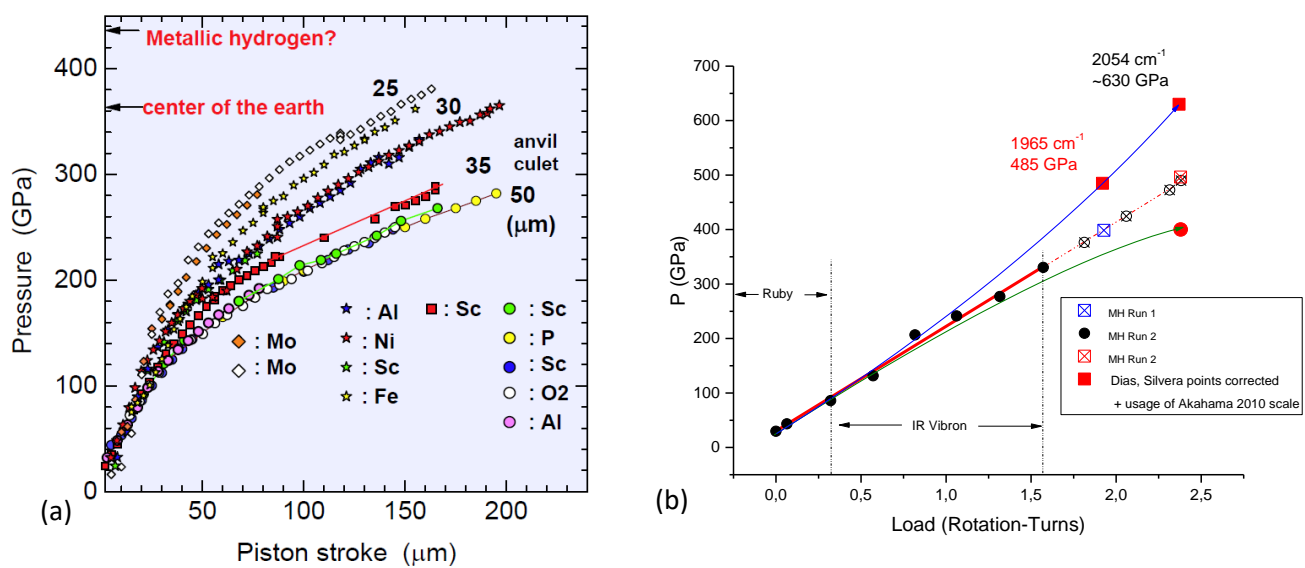

Fig. 1. Determination of pressure. Left - Pressure/load dependence for diamond anvils with different diameter of culets (Akahama 2007). Right - The linear loading curve is taken from Ref. (Dias and Silvera 2017). Red square points - the pressure was determined by us from the step at the original Raman spectra (SM Fig. 3) and using Akahama 2010 pressure scale (see discussion in the text). Red circle is a pressure determined with Akahama 2010 scale and from the middle of the smooth step in the spectra.

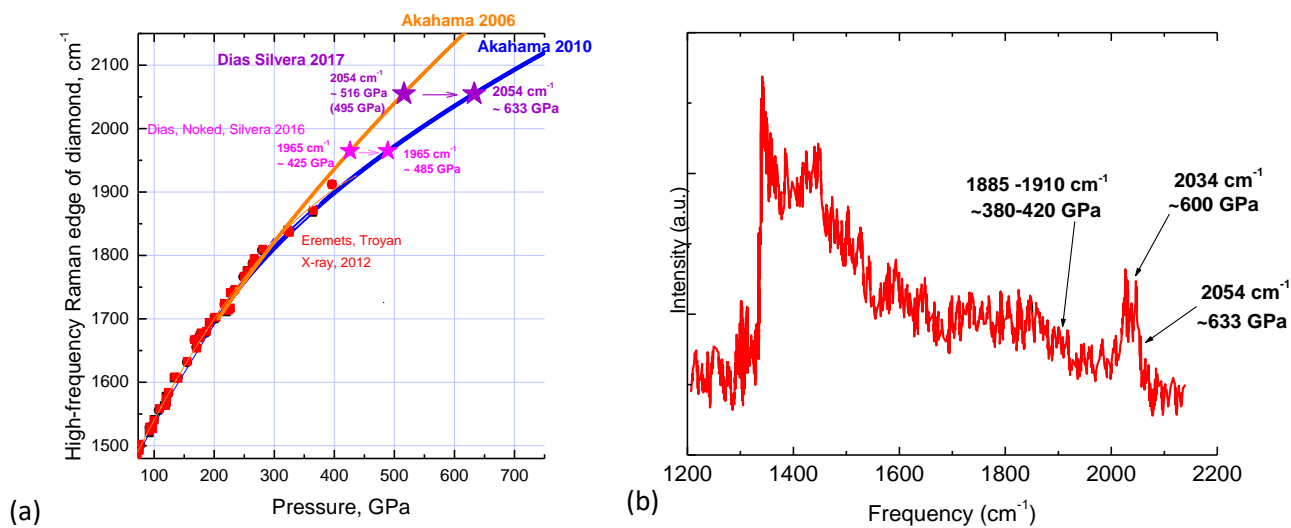

Fig. 2. Pressure determination from the shift of Raman spectra of the stressed diamond anvils. (a) Pressure scales: From Ref. (Akahama and Kawamura 2006) - orange line, blue line scale from Ref. (Akahama and Kawamura 2010). Red point our scale (Eremets 2003) which was extended to 400 GPa by measuring EOS of gold. (b) Raman spectrum taken from Ref. (Dias and Silvera 2017). The pressures were assigned by us to the characteristic points of the spectrum (steps) as required by the (Akahama and Kawamura 2010) pressure scale.

The diamond pressure scale is well established by calibration against the equation of state (EOS) of metals up to 410 GPa (Eremets 2003, Akahama and Kawamura 2004, Sun, Ruoff et al. 2005, Akahama and Kawamura 2010, Eremets and Troyan 2012). Initially, a linear dependence of pressure on the Raman shift was proposed (Akahama and Kawamura 2006) but it is valid only up to 300 GPa. At higher pressures a nonlinear scale (Akahama and Kawamura 2010) should be used. We confirmed this by measuring (EOS) of gold up to 400 GPa (Eremets and Troyan 2012). This, well established scale (Akahama and Kawamura 2010), gives a pressure of 633 GPa from the position of the middle of the step (2054 cm$^{-1}$) shown by arrow at the spectrum of Fig. S2 (Dias and Silvera 2017), see also Fig. 2 in the present text. Erroneously the authors assigned 2034 cm$^{-1}$ to the step, which gives smaller value of pressure. Moreover, DS use the old linear scale (Akahama and Kawamura 2006) because ".. our experience with our DAC using the pressure vs. load scale motivated us to use the more conservative linear scale of Akahama and Kawamura". Note, that DS used the nonlinear scale in their previous work (Dias, Noked et al. 2016). With this linear scale DS lowered pressure from 633 GPa to the announced 495 GPa, which had the effect of putting it closer to their linear pressure-load "scale", and make the value of pressure apparently more acceptable. Nevertheless, the achievement of 495 GPa (in fact 633 GPa) with culets of 30-35 micrometer and single-beveled diamond anvils looks very surprising, it is far above the current record of pressure in DAC 400-450 GPa reached with culets of 15-20 micrometers (smaller culets give a definite advantage, see Fig. 1). The authors argue that they used specially prepared diamond anvils where surface defects were removed by etching of the diamonds after polishing, and illuminated the sample with precaution. But all this is known, we use the same techniques in past years. The etching (removal of defects) indeed improves the reproducibility in achieving pressures of about 400 GPa but doesn't give a great addition to the achievable pressure. The loading curve of DS gives clear indication that the determination of pressure is most likely wrong and greatly overestimated. The correctly defined pressure of 633 GPa from the spectrum in Fig. 2 makes the loading curve superlinear (Fig. 1b) which is physically impossible.

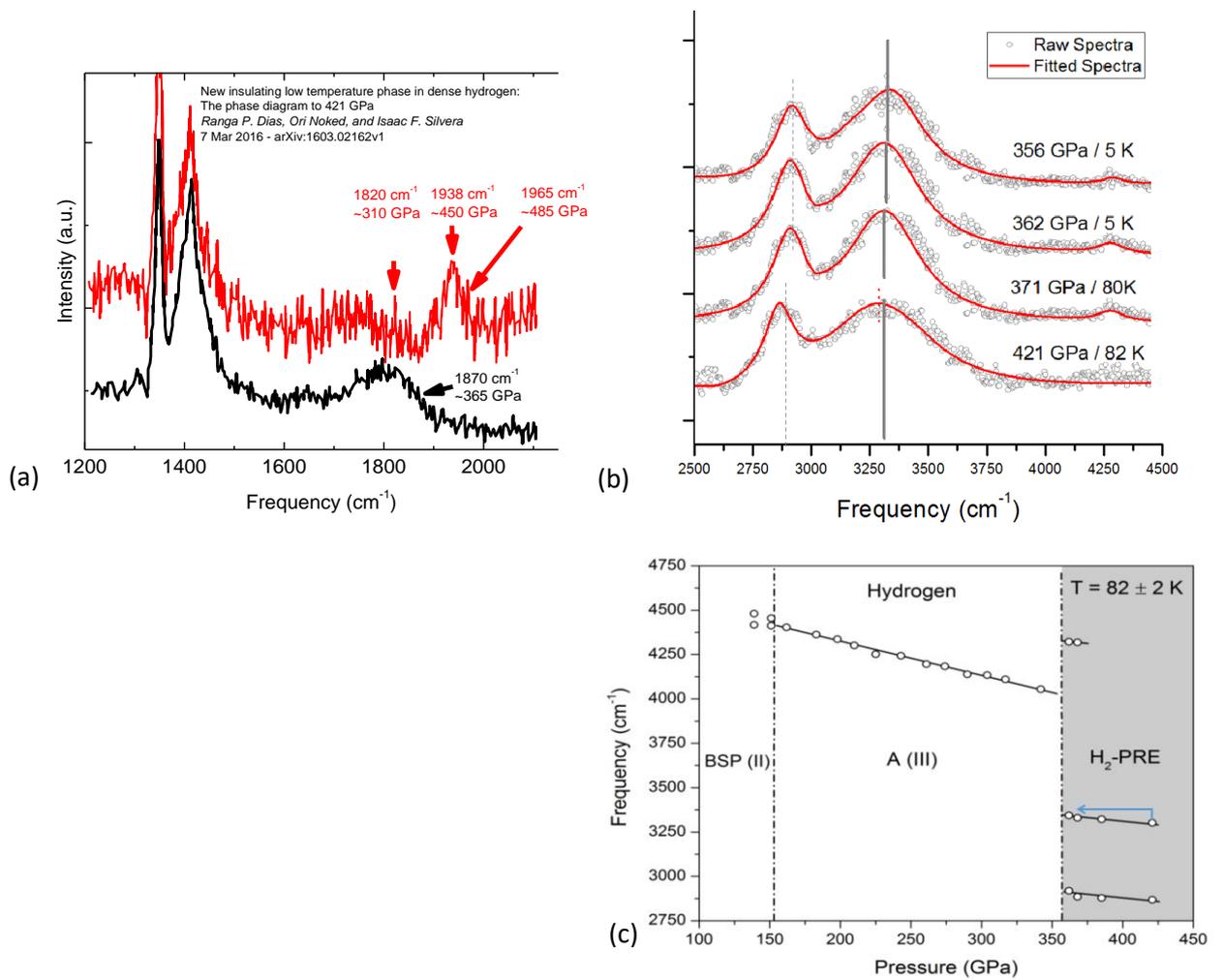

Fig. 3. Determination of pressure in Ref. (Dias, Noked et al. 2016). (a) Comparison of the Raman spectra at subsequent loadings. (b) Changes of the infrared spectra. (c) Pressure dependence of the IR spectra. See the text for explanations.

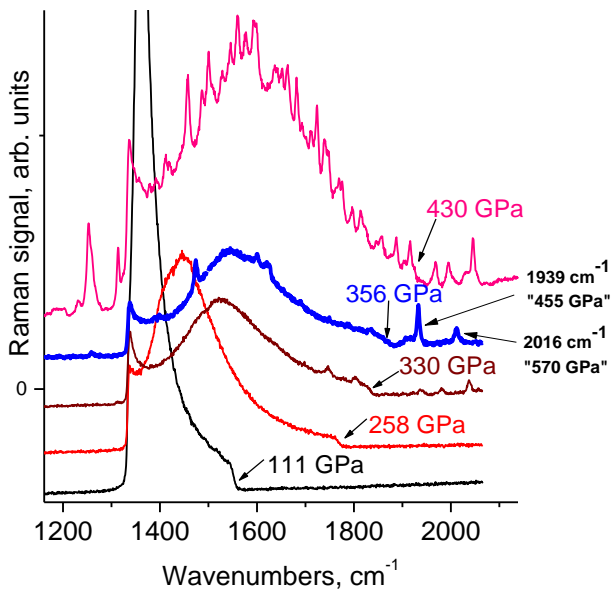

Fig. 4. Evolution of Raman spectra from stressed diamond anvils in some of our experiments. At pressures above ~330 GPa spurious peaks appear which likely are luminescence originated from defects in the stressed anvils. See also Ref. (Akahama and Kawamura 2010). For instance, the spectrum corresponding to 356 GPa (blue line) contain peaks which can be erroneously indicate higher pressures.

The determination of pressure in the previous work of the authors (Dias, Noked et al. 2016), see Fig. 3a (see also Ref (Dias, Noked et al. 2016), Fig. S2) gives additional support for suspicion that the pressure was wrongly determined. The lower spectrum has a step at 1880 cm$^{-1}$, which gives pressure of ~380 GPa according to the scale of (Akahama and Kawamura 2010). The upper spectrum corresponds to the next loading. Here again, mistakenly, a top of the peak was taken for the pressure determination instead of the required middle of the step. The middle of the step gives a pressure of 485 GPa. In this case, the corresponding frequency of the vibron (Dias, Noked et al. 2016), Fig. 2a) should significantly decrease according to the observed pressure tendency, but in fact it almost did not change as is clearly evident from the spectra (Fig. 3b). Incorrectly, the position of the peak was put at the *lower* frequency in the summary plot (Fig. 3c copy of Ref (Dias, Noked et al. 2016)). On closer examination, the fit of the raw spectra (Fig 3b copy of Ref (Dias, Noked et al. 2016)) seems not to be correct but should be shifted to higher frequencies, thus the position of this peak does not move with pressure either. The unchanged position of the vibron suggests that the pressure after loading in fact does not change and remains at about 365 GPa, but the very high pressure of 485 GPa was mistakably assigned to the peak likely originated, not from Raman signal, but from a luminescence line.

Random peaks often appear at the highest pressures, before the diamonds break. This was documented in Ref. (Akahama and Kawamura 2010), we also observed them in many experiments (Fig. 4). These peaks likely originate from defects which develop prior the failure. They may lead to misleading interpretation of the spectrum, and wrong determination of pressure. Normally it is difficult to make a rough mistake in the pressure when the Raman spectra are tracked with gradual increase of load (Fig. 4). However, if there is only one spectrum (as measured in Ref (Dias and Silvera 2016)), the interpretation can be ambiguous.

We guess that the peaks in Ref. (Dias and Silvera 2017) (Fig. S2) and (Dias, Noked et al. 2016) Fig. S2, can also be spurious. The pressure in Ref. (Dias and Silvera 2017) should be determined, not from this peak, but from the smooth step at lower frequencies, *i. e.* from the 1885 -1910 cm$^{-1}$ point in Fig. 2b. In this case the pressure is ~380-420 GPa according to the (Akahama and Kawamura 2010) scale. These pressures are still very high but probably might be achieved in Ref. (Dias and Silvera 2017).

If the real pressure in the DS experiment is in the range of about ~380 GPa, this would mean that the reflecting sample actually relates to the transition which we found at ~360-380 GPa (M. I. Eremets, Troyan et al. 2016). We observed a transition to a new low temperature conductive, possibly metallic, phase as followed from the drop of resistance and the disappearance of Raman spectra. However, we did not claim metallic hydrogen on the basis of a single experiment. We could not reach these pressures and reproduce the experiment for more than three years until last year when we reproduced the electrical measurements in three experiments and disappearance of the Raman spectra at low frequencies on cooling below 200 K. Still, we believe that more experiments should be performed for definite conclusion of the state of hydrogen. In particular, reflection alone can be confusing – we observed reflecting hydrogen starting from ~ 350 GPa and low temperatures (Fig. 5a), but the sample was not metallic – it had Raman signal characteristic for phase III. Our observation of the reflective sample at P~350 GPa can be another indication that DS actually were below 400 GPa as they stopped loading at the first appearance of the reflectivity.

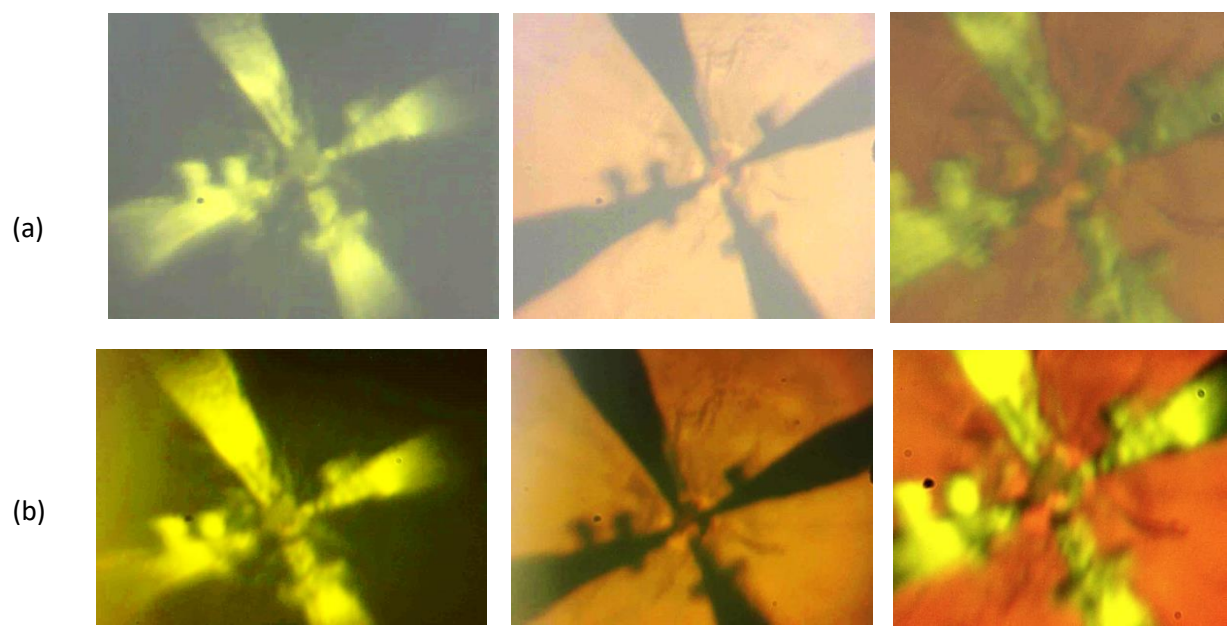

(a)

(b)

Fig. 5. Sample of hydrogen at 337 GPa (upper row) and 380 GPa (below) and 170 K: Photographs were taken in reflected, transmission and combined reflected/transmission light. The sample of hydrogen is in the center, it has nearly round shape. Four gold electrical leads touch the sample. The sample, not culet reflects as it is clear from the right enlarged photographs taken in the combined transmission and reflection illumination.

**Reflectance measurements**

First, there are questions about the experimental arrangement. The author should prove that the observed reflection is from the hydrogen sample and experimentally exclude a possibility of reflection from the layer of alumina at the surface of diamond anvil. Amorphous alumina has a band of ~3 eV (Århammar, et al. 2011) and it might be closed at very high pressures. Other materials for coating should be used to exclude a possibility of reflection from alumina which might react with hydrogen at very high pressures. Reaction of hydrogen with the surface of diamond at megabar pressures is also possible (Liu, Naumov et al. 2016) and should be taken into consideration.

The only direct way to establish a metallic state is to measure the electrical conductivity of the sample down to the lowest temperatures: metal has free electrons and conducts to the lowest temperatures in both normal and superconducting states. In the absence of electrical measurements, the authors measured reflectivity and used the Drude model which describes the spectrum of reflectivity of free electrons. This indirect method requires measurements of reflectivity in the entire spectral range. The authors measured reflectivity only in four points in visible part of the spectrum. First, it is not clear how these primarily data were determined: what was the reference surface (back of diamond anvil?), was the incident beam on the axis of the cell or at some angle? The culet is apparently not flat, but it is convex. This effect is not considered but it can introduce significant error for the reflectance measurements. Visually, the central part of a convex mirror should be looked brighter. Thus comparing reflectance of the sample and gasket become pointlessly.

The stressed anvils also absorb light: the band gap of diamond decreases significantly with uniaxial stresses and the reflection data should be corrected for this effect. DS used absorption spectra of the stressed anvils from Vohra's work (Vohra 1991). This is a very questionable procedure as the absorption spectra depend on particular stresses in the diamond anvil and are determined by the geometry of the anvils, the material of the gasket and other uncontrolled parameters; they strongly differ from DAC to DAC (Fig. 5). Additional large uncertainty is produced by the extrapolation of the data (Y. K. Vohra 1992) to much higher pressures.

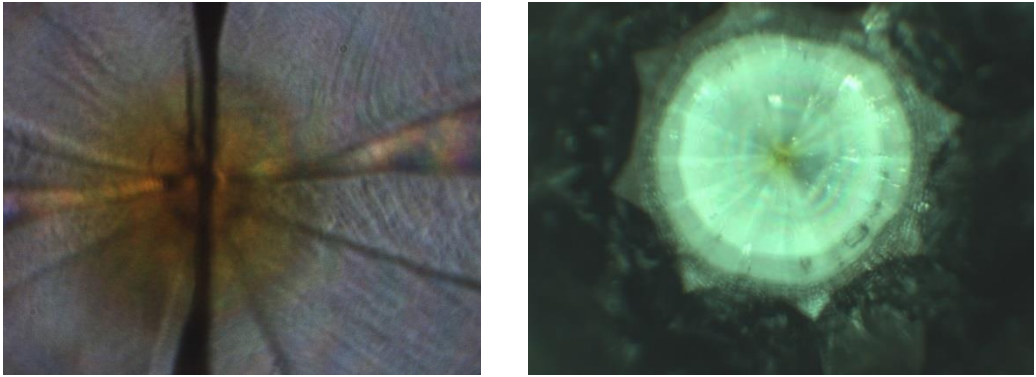

Fig. 6. Photographs of anvils in the combined transmission and reflection illumination at similar pressures of 370 GPa and 380 Gpa. The left photograph represent arrangement for electrical measurements described in Ref [18]. The very stressed tip of diamond is brown. This color is typical for high stressed diamond before failure. In some runs (right) the anvils are only yellowish at 380 GPa.

The procedure of the correction seems to be straightforward but we was not able to reproduce it. In particular, the way how the two-pass optical density of stressed diamond at high pressure was taken from the valuable work (Vohra 1991). It is not easy available and therefore we present here some plots (Fig. 7). In particular, it is clear that DS have optical density >3 for 3.06 eV (Fig. 7). But there is no such number in Vohra measurements of optical density for type IIa anvil, while the optical pass in Vohra and DS measurements is nearly the same ~5 mm.

We replotted the optical density dependence on pressure for different energy from original work of Vohra (Fig. 7) and the difference with the DS work is significant. This difference is unclear.

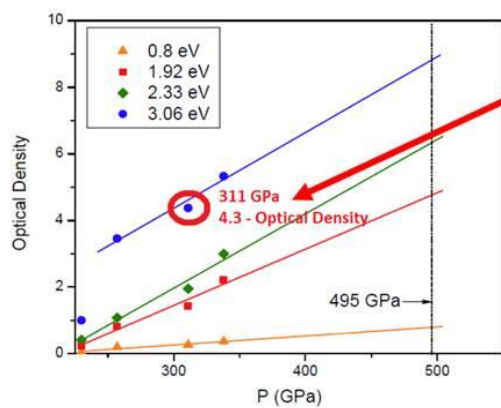
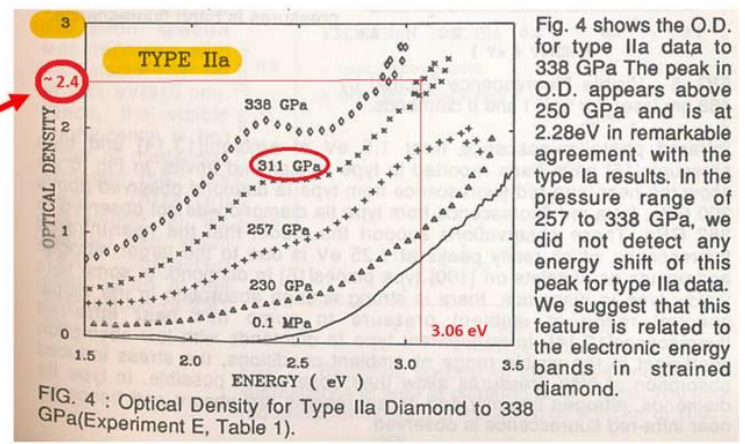

(a)

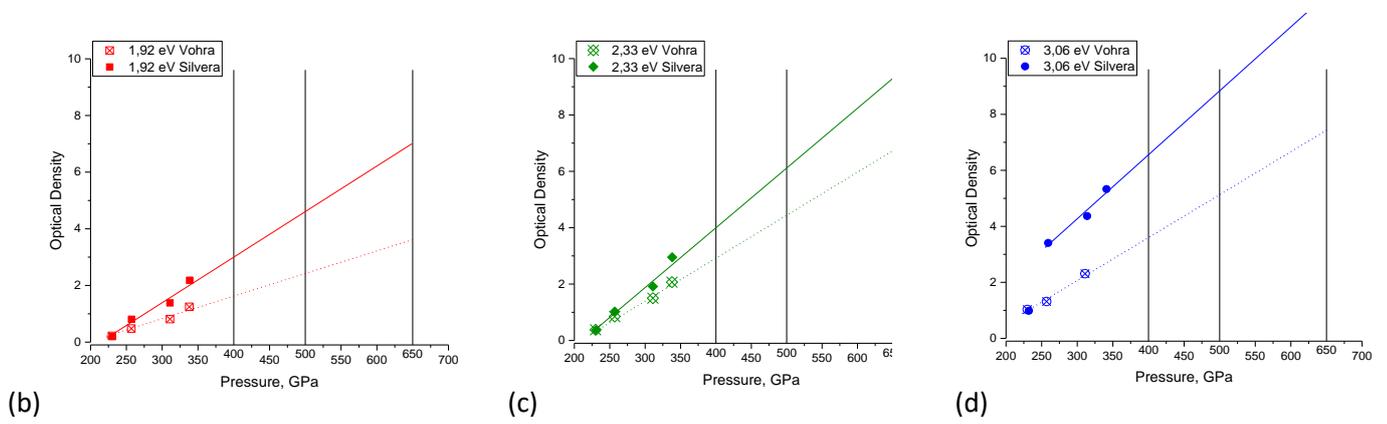

Fig. 7. The two-pass optical density of stressed diamond at high pressure for different spectral range.

The extrapolation of the optical density to 495 GPa give unrealistically high values. These values should be even higher in extrapolation to 633 GPa. For instance, for the 3.06 laser line the optical density would be 8! (i.e. attenuation is 10-8), for extrapolation to 495 GPa or even ~12 for 630 GPa – impossibly high. Nevertheless, these unrealistic data were used to correct the reflection data. As a result, the reflection spectrum (four points) became nearly flat, and it was used for fits with a Drude model (two fitting parameters). This very questionable procedure led to an important conclusion on high concentration of free electron consistent with atomic hydrogen! See also comment arXiv:1702.04246.

**Conclusions**

To conclude, we consider that the claim of the achievement of metallic hydrogen (Dias and Silvera 2016, Dias and Silvera 2017) was not established.

- The pressure measurements presented are highly unreliable and the pressure can be estimated to be either 630 GPa or 380 - 420 GPa. We showed that using a correct calibration scale step of the peak of the Raman spectra yields a pressure of 630 GPa. We argue that there is no mystery about these "record pressures", and more likely actual pressure in DS experiment did not exceed ~380-420 GPa.

- The calculations of reflectance and concentration of free electrons are based on a poorly estimated pressure and the optical density of diamond. Both of these values have too big (up to 100%) uncertainty to make the estimation of free electrons concentration trustworthy.

- The visual observation of the enhanced reflectivity cannot be considered as evidence. The photo of a reflecting sample is the most catching part of Dias and Silvera work but hydrogen in nonmetallic semiconducting state also reflects well at pressures above ~300 GPa and low temperatures as we observed in our experiments.

Likely actual pressure in DS experiment did not exceed ~360-400 GPa. In this case the observation of the reflective hydrogen can support our work (M. I. Eremets, Troyan et al. 2016) on a new phase at pressures above 360-380 GPa where we found the reproducible disappearance of Raman peaks, and this phase is conductive, also reproducible, with a mixed metallic-semiconducting temperature behavior.

Finally, we acknowledge contribution of P. P. Kong and H. Wang in the mentioned results on hydrogen, in particular, in Fig. 5. The full scope of results will be published elsewhere.

___

**References**

Akahama, Y. (2007). "Diamond Anvil Raman Gauge in Multimegabar Range." Workshop on Pressure scale Jan. 26-28 2007 Geophysical Lab, CIW.

Akahama, Y. and H. Kawamura (2004). "High-pressure Raman spectroscopy of diamond anvils to 250 GPa: Method for pressure determination in the multimegabar pressure range." J. Appl. Phys. **96**: 3748-3751.

Akahama, Y. and H. Kawamura (2006). "Pressure calibration of diamond anvil Raman gauge to 310 GPa." J. Appl. Phys. **100**: 043516

Akahama, Y. and H. Kawamura (2010). "Pressure calibration of diamond anvil Raman gauge to 410 GPa." J. Physics. C(215, 012195).

Århammar, C., A. P., N. and e. al (2011). "Unveiling the complex electronic structure of amorphous metal oxides." PNAS **108**: 6355-6360.

Dias, R., O. Noked and I. F. Silvera (2016). "New low temperature phase in dense hydrogen: The phase diagram to 421 GPa." arXiv:1603.02162.

Dias, R. and I. F. Silvera (2016). "Observation of the Wigner-Huntington Transition to Solid Metallic Hydrogen." arXiv:1610.01634.

Dias, R. P. and I. F. Silvera (2017). "Observation of the Wigner-Huntington transition to metallic hydrogen." Science: 10.1126/science.aal1579.

Eremets, M. I. (2003). "Megabar high-pressure cells for Raman measurements." J. Raman Spectroscopy **34**: 515–518.

Eremets, M. I. and I. A. Troyan (2012). "Diamond edge pressiure scale calibrate to 400 GPa with EOS of gold." unpublished.

Liu, H., I. I. Naumov and R. J. Hemley (2016). "Dense Hydrocarbon Structures at Megabar Pressures." J. Phys. Chem. Lett. **7**: 4218–4222.

M. I. Eremets, I. A. Troyan and A. P. Drozdov (2016). "Low temperature phase diagram of hydrogen at pressures up to 380 GPa. A possible metallic phase at 360 GPa and 200 K." arXiv:1601.04479.

Sun, L., A. L. Ruoff and G. Stupian (2005). "Convenient optical pressure gauge for multimegabar pressures calibrated to 300 GPa." Appl. Phys. Lett. **86**: 014103.

Vohra, Y. K. (1991). Spectroscopic studies of diamond anvil under extreme static stresses. AIRAPT: Recent Trends in High Pressure Research, Bangalore, India, Oxford Press, New Delhi, India.

Y. K. Vohra (1992). Spectroscopic studies on diamond anvil under extreme static pressures The XIII AIRAPT International Conference, Recent Trends in High Pressure Research. A. K. Singh. Bangalore, Oxford Press, New Delhi, India, 1992**:** 349-358.

Zha, C.-S., Z. Liu and R. J. Hemley (2012). "Synchrotron Infrared Measurements of Dense Hydrogen to 360 GPa." Phys. Rev. Lett. **108** 146402.